# Diffuson-driven Ultralow Thermal Conductivity in Amorphous Nb$_2$O$_5$ Thin Films

Zhe Cheng[1], Alex Weidenbach[2], Tianli Feng[3,4], M. Brooks Tellekamp[2], Sebastian Howard[5], Matthew J. Wahila[5], Bill Zivasatienraj[2], Brian Foley[1], Sokrates T. Pantelides[3,4], Louis F.J. Piper[5,6], William Doolittle[2], Samuel Graham[1,7,*]

[1] George W. Woodruff School of Mechanical Engineering, Georgia Institute of Technology, Atlanta, Georgia 30332, USA

[2] School of Electrical and Computer Engineering, Georgia Institute of Technology, Atlanta, GA, 30332, USA

[3] Department of Physics and Astronomy and Department of Electrical Engineering and Computer Science, Vanderbilt University, Nashville, Tennessee 37235, USA

[4] Materials Science and Technology Division, Oak Ridge National Laboratory, Oak Ridge, Tennessee 37831, USA

[5] Department of Physics, Applied Physics and Astronomy, Binghamton University, Binghamton, New York 13902, USA

[6] Materials Science & Engineering, Binghamton University, Binghamton, New York 13902, USA

[7] School of Materials Science and Engineering, Georgia Institute of Technology, Atlanta, Georgia 30332, USA

* Corresponding Author: sgraham@gatech.edu

# Abstract

Niobium pentoxide ($Nb_2O_5$) has been extensively reported for applications of electrochemical energy storage, memristors, solar cells, light emitting diodes (LEDs), and electrochromic devices. The thermal properties of $Nb_2O_5$ play a critical role in device performance of these applications. However, very few studies on the thermal properties of $Nb_2O_5$ have been reported and a fundamental understanding of heat transport in $Nb_2O_5$ is still lacking. The present work closes this gap and provides the first study of thermal conductivity of amorphous $Nb_2O_5$ thin films. Ultralow thermal conductivity is observed without any size effect in films as thin as 48 nm, which indicates that propagons contribute negligibly to the thermal conductivity and that the thermal transport is dominated by diffusons. Density-function-theory (DFT) simulations combined with a diffuson-mediated minimum-thermal-conductivity model confirms this finding. Additionally, the measured thermal conductivity is lower than the amorphous limit (Cahill model), which proves that the diffuson model works better than the Cahill model to describe the thermal conduction mechanism in the amorphous $Nb_2O_5$ thin films. Additionally, the thermal conductivity does not change significantly with oxygen vacancy concentration. This stable and low thermal conductivity facilitates excellent performance for applications such as memristors.

## Introduction

Niobium pentoxide ($Nb_2O_5$) has been extensively used for applications of electrochemical energy storage[1-5], solar cells and light emitting diodes (LEDs)[6-9], electrochromic devices[10, 11], and memristors[12-14]. The thermal management in electrochemical energy-storage devices such as batteries and supercapacitors, and solar cells and LEDs, is critical for device performance, reliability, and safety.[15-17] For electrochromic applications, such as smart windows, thermal insulation plays an important role in building energy savings.[10] For memristors, the electrical properties are directly related to the local temperature field and thermal insulation is a key for reducing device power consumption.[18-20] However, very few experimental studies on the thermal properties of $Nb_2O_5$ have been performed and a fundamental understanding of heat transport in this material, especially the amorphous state, is still lacking. In amorphous materials, heat is carried by local vibrations (locons), standing non-localized vibrations (diffusons), and propagating vibrations (propagons),[21] while the phonon gas model which was based on vibrations in an infinitely large and pure crystal is still extensively used in applications.[11, 22, 23] Some amorphous materials like amorphous silicon show a size effect in their thermal conductivity because propagons are a major contributor, while some others like amorphous $SiO_2$ do not as their thermal transport is dominated by diffusons.[24, 25] It is interesting and important to explore how heat carriers contribute to thermal conductivity of $Nb_2O_5$ for both fundamental understanding and practical applications.[13, 17, 19, 22, 23, 26]

In this work, we report the thermal conductivity of amorphous $Nb_2O_5$ thin films for the first time. Time-domain thermoreflectance (TDTR) is used to measure the thermal conductivity of these thin films and DFT simulations combined with a diffuson-mediated minimum-thermal-conductivity model is used to understand the measured thermal conductivity. Additionally, the effects of oxygen-vacancy concentration, thickness, and temperature on the thermal conductivity of amorphous $Nb_2O_5$ are studied.

## Results and discussions

### Sample structure

The $Nb_2O_5$ thin films were grown on sapphire substrates by room-temperature reactive sputtering. The oxygen-vacancy concentration was varied within the samples by modifying the Nb target power with respect to a constant oxygen pressure. As such, the films are henceforth referred to as $Nb_2O_{5-x}$ where x represents the deviation from ideal stoichiometric oxygen content (oxygen vacancies). The thin films were characterized by x-ray diffraction (XRD) to confirm that they are amorphous. X-ray photoelectron spectroscopy (XPS) was used to determine the ratio of Nb and O atoms in the films.[27] The Nb:O ratios of films with different growth powers are shown in Table 1. Each reported ratio is an average of 17 measurements taken on different sample spots. The given error bars are the statistical variations of the 17 measured spots for each sample. The surface oxygen component in the O 1s core-level was subtracted. All the grown films are oxygen deficient compared to stoichiometry (common for most amorphous metal oxides) and the Nb:O ratio increases slightly with the growth power.

Table 1. Nb:O ratios of thin films with different growth powers.

| Growth power | 30 W | 40 W | 50 W |
|---|---|---|---|
| Nb:O ratio | 0.479 (+0.040/-0.066) | 0.482 (+0.022/-0.080) | 0.497 (+0.054/-0.038) |

X-ray reflectometry (XRR) is used to measure the film thickness for thin films (thinner than 200 nm) while profilometer (Tencor P15 Profilometer) is used to measure the film thickness for thicker films. A layer of Al (~80 nm) is deposited on the $Nb_2O_{5-x}$ films as transducer for TDTR measurements.

**Phonon properties of single crystal $Nb_2O_5$**

DFT calculations have been performed for single-crystal $Nb_2O_5$ with space group C2/m (seven atoms per unit cell). The phonon dispersion relation and phonon DOS are shown in Figure 1(a-b). The asymmetric complex crystal structure leads to a large number of optical phonon branches. Optical phonons usually contribute a very small amount to thermal conductivity, but scatter with acoustic phonons.[28] These optical phonons scatter extensively with acoustic phonons, leading to a small phonon-phonon scattering relaxation time and subsequently low thermal conductivity. There is no phonon bandgap between the low frequency optical phonons (3-12 THz) and acoustic phonons, which results in extensive optical phonon-acoustic phonon interaction that reduces thermal conductivity. Taking the (Γ-X) direction for example, two phonon bandgaps (15-18 THz and 20-24 THz) are observed. These high frequency optical phonons are less likely to interact with acoustic phonons than the low frequency optical phonons. Correspondingly, there are two DOS valleys at these frequency ranges in Figure 1(b). It is noteworthy that niobium contributes very little to the high-frequency DOS.

The overall phonon DOS is projected to Nb and O species in Figure 1(b). The total area enclosed by the pDOS of oxygen is 2.5 times of that of niobium since oxygen has 2.5 times of the stoichiometric amount. For frequencies larger than 15 THz, the optical phonon branches come from oxygen vibrations because of the small atomic mass of oxygen. In the acoustic phonon frequency range (0-3 THz), the DOS contributed by niobium is larger than that by oxygen. Heavy atoms dominate in low-frequency acoustic vibrations and contribute to thermal conductivity. For high frequencies (>13 THz), we can see from Figure 1(b), the DOS contributed by niobium is almost zero. According to the phonon dispersion relation in Figure 1(a), we determine the average phonon group velocity of $Nb_2O_5$ to be approximately about 5000 m/s, which is very close to the sound velocity obtained from the picosecond acoustic method in TDTR measurements.

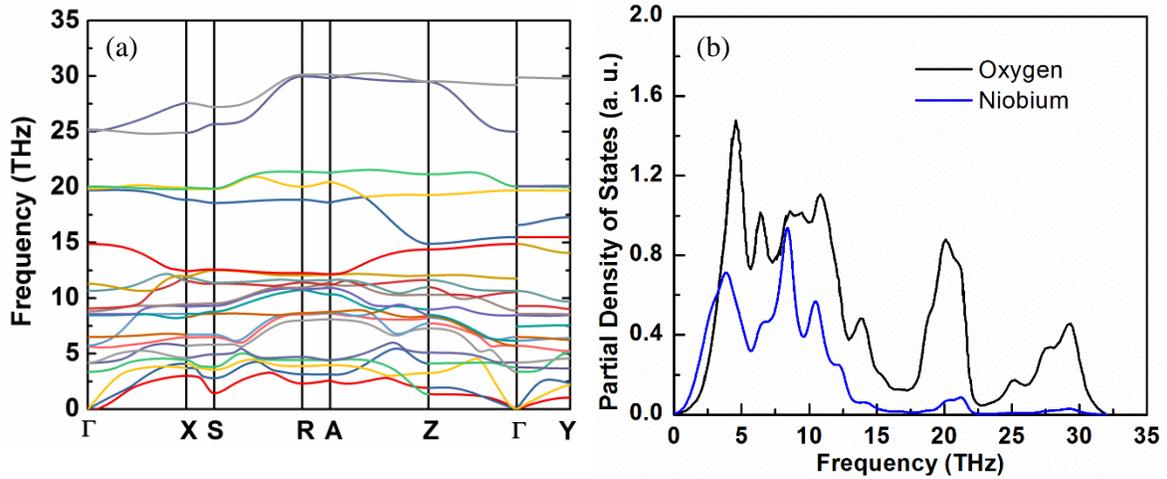

Figure 1. DFT calculated (a) phonon dispersion relation (b) partial phonon density of states of $Nb_2O_5$.

The temperature dependent Helmholtz free energy, entropy, and heat capacity of single crystal $Nb_2O_5$ have also been calculated and reported in Figure 2. The relation between Helmholtz free energy ($F$) and entropy ($S$) is $S = - \partial F/\partial T$. The Helmholtz free energy and entropy are of great significance in the calculation of thermodynamical properties, such as the thermal expansion coefficient. The heat capacity is compared with literature values and excellent agreement is achieved.[29] As the temperature increases, more high-energy phonon modes are excited, resulting in an increasing heat capacity. At high temperatures (>500 K), all the phonons modes are fully excited, resulting in a constant heat capacity. The heat capacity values shown here are used in the TDTR data fittings.

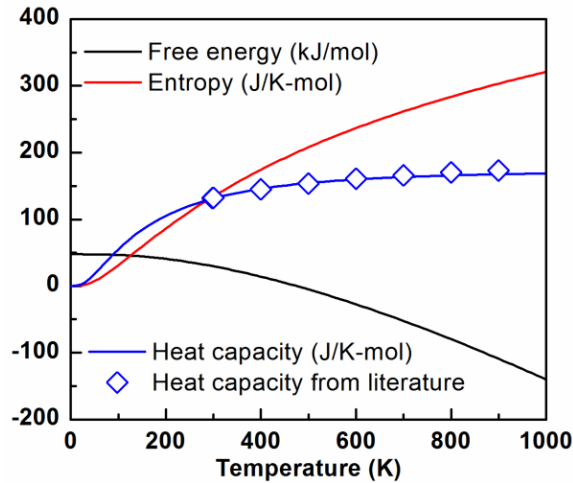

Figure 2. DFT calculated heat capacity, Helmholtz free energy, and entropy of single crystal $Nb_2O_5$. The calculated heat capacity is compared with measured values in the literature.[29]

**Effects of thickness and growth power on thermal conductivity**

The effects of thickness and growth power on the measured thermal conductivity of $Nb_2O_{5-x}$ films are shown in Figure 2. The thermal conductivity of $Nb_2O_{5-x}$ films with thicknesses as low as 48 nm was measured. The thermal conductivity is around 1 W/m-K and does not change with thickness (no size effect). The present work serves as a benchmark for the thermal properties of amorphous $Nb_2O_{5-x}$ films, which are needed for multi-physics modelling in memristors and other applications.[20, 30, 31] Moreover, the thickness-dependent thermal conductivity data show that $Nb_2O_{5-x}$ is different from amorphous silicon or SiC in which propagons contribute a large fraction of the thermal conductivity.[24, 32-34] In $Nb_2O_{5-x}$, no thickness dependence is observed, which indicates that propagons contribute negligibly to the thermal conductivity of these films.

Therefore, we conclude that diffusons dominate the thermal transport in amorphous $Nb_2O_{5-x}$ thin films.

Similar to thickness, the data show that growth power does not affect the thermal conductivity of $Nb_2O_{5-x}$ films either. The growth power is correlated to oxygen vacancy concentration in the films, as shown in Table 1. The growth power of the Nb target determines the amount of released Nb from the target while the oxygen pressure is constant. The higher the grown power is, the larger amount of Nb is released. As a result, the Nb:O ratios, subsequently oxygen vacancy concentrations, are varied with grown powers. The samples grown with higher growth powers have higher oxygen vacancy concentrations. However, these vacancies do not affect the thermal conductivity because the lattice vibrations in the films are localized (locons) or short-range propagated (diffusons). Locons usually contribute very little to thermal conductivity because they do not propagate and transport heat.[25, 33] Diffusons dominate heat transport in these films. Unlike phonons in crystalline materials, both locons and diffusons are not sensitive to point defects like oxygen vacancies in the films because the amorphous films themselves are in disorder.[35]

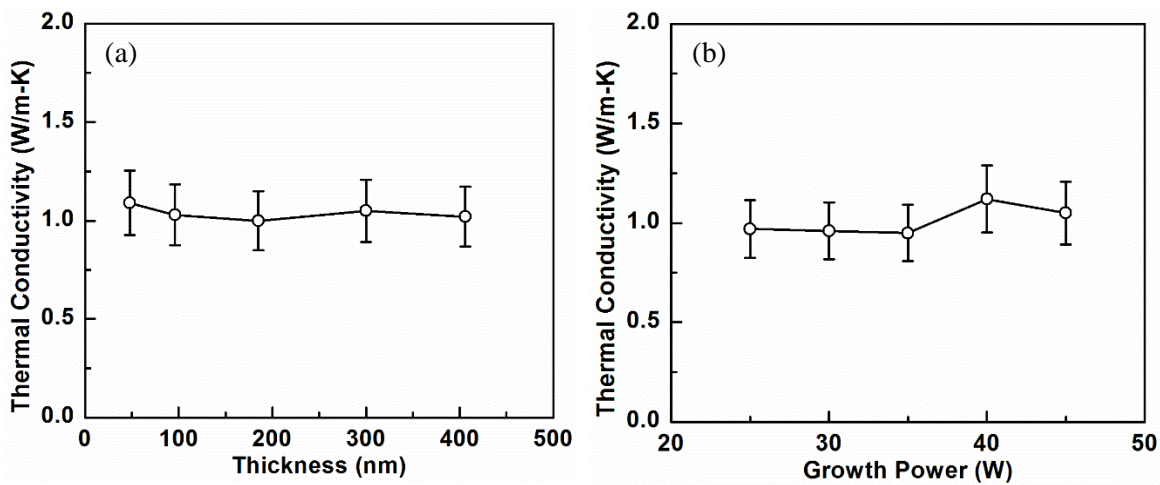

Figure 3. (a) the effect of film thickness on thermal conductivity. (b) the effect of growth power on thermal conductivity.

**Diffuson-driven ultralow thermal conductivity**

The temperature-dependent thermal conductivity of $Nb_2O_{5-x}$ films has also been measured by TDTR to explore the thermal conduction mechanism. In amorphous materials, the concept of a phonon does not hold. The definition of phonon group velocity in the kinetic-theory-based phonon thermal conductivity $\kappa = C_v l v/3$ is not applicable either. By assuming the mean-free-paths of the Debye-like, heat-carrying oscillations are half of the corresponding wavelengths, Cahill developed a minimum thermal conductivity model[36-38]:

$$\kappa_{Cahill} = \left(\frac{\pi}{6}\right)^{1/3} k_B n^{2/3} \sum_{i=1}^{3} v_i \left(\frac{T}{\Theta_i}\right)^2 \int_0^{\Theta_i/T} \frac{x^3 e^x}{(e^x-1)^2} dx \ , \qquad (1)$$

where $k_B$ is the Boltzmann constant, $v_i$ is the sound velocity of polarization $i$, $n$ is the atomic density, and $\Theta_i = v_i(h/2\pi k_B)(6\pi^2 n)^{1/3}$ is the cutoff frequency expressed as temperature unit (similar to Debye temperature for crystalline materials). $h$ is the Planck constant. The calculated values of this model are referred to as the amorphous limit.

The thermal conductivity can also be expressed as $\kappa = \int_0^\infty g(\omega) C(\omega) D(\omega) d\omega$ to avoid the definition of phonon group velocity. Here, $g(\omega)$ is the density of states at frequency $\omega$; $C(\omega)$ is the heat capacity; $D(\omega)$ is the thermal diffusivity. Recently, by introducing a diffuson diffusivity, Agne *et. al.* developed another minimum thermal conductivity model for diffuson-mediated thermal transport[39]:

$$\kappa_{diffuson} = \frac{n^{1/3}k_B}{\pi} \int_0^\infty \left(\frac{g(\omega)}{3n}\right) \left(\frac{\hbar\omega}{2\pi Tk_B}\right)^2 \frac{e^{\frac{\hbar\omega}{2\pi Tk_B}}}{\left(e^{\frac{\hbar\omega}{2\pi Tk_B}} - 1\right)^2} \omega d\omega \,. \qquad (2)$$

The only input in this model is the density of states $g(\omega)$, which is obtained from DFT calculations, as shown in Figure 4(a). The direct calculation of the density of states of an amorphous material is not practical in DFT. We, therefore, use the single-crystal phase to approximate the amorphous material's DOS since the two are usually similar.[40] By inserting the DFT-calculated DOS into Equation (2), the minimum thermal conductivity of diffuson-mediated thermal transport is obtained.

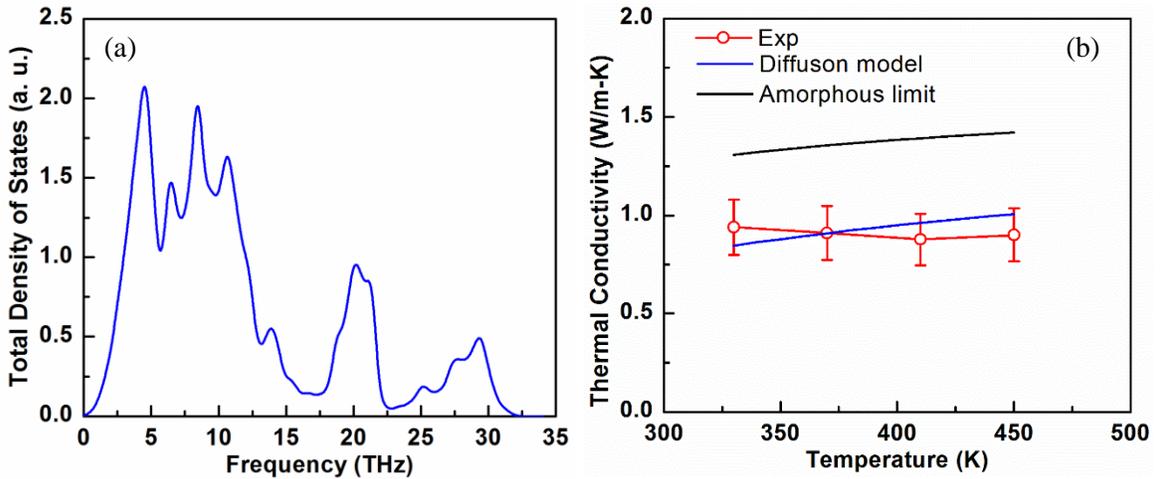

Figure 4. (a) total density of states from DFT calculation. (b) temperature dependent thermal conductivity. Two minimum thermal conductivity models are calculated to compare with the experimental values.

As shown in Figure 4(b), we observe a weak temperature dependence of the measured thermal conductivity of the $Nb_2O_{5-x}$ films. The measured thermal conductivity is lower than the amorphous limit (Cahill model), while it agrees well with the diffuson model. The small discrepancy may arise from the small difference between the density of states of amorphous and single-crystal phases. This result agrees well with the conclusion in Sec. 2.3 that diffusons dominate the thermal transport. For memristor applications, the stable and ultralow thermal conductivity facilitates improved device performance by lowering the threshold voltage, threshold current, and device power consumption.[20]

## Conclusions

Niobium pentoxide ($Nb_2O_5$) has been extensively studied for applications of electrochemical energy storage, memristors, solar cells and light emitting diodes (LED), and electrochromic devices. The thermal properties of $Nb_2O_5$ play a critical role in device performance of these applications. However, very few studies on thermal properties of $Nb_2O_5$ have been reported and a fundamental understanding of heat transport in $Nb_2O_5$ is still lacking. In the present work, we reported for the first time the thermal conductivity of amorphous $Nb_2O_5$ thin films with TDTR and obtained an ultralow thermal conductivity of 1 W/m-K. No size effect is observed in the measured thermal conductivity with film thickness as thin as 48 nm, indicating that propagons contribute negligibly to the thermal conductivity and the thermal transport is dominated by diffusons. DFT calculations combined with a diffuson-mediated minimum thermal conductivity model confirms this finding. The DFT-calculated heat capacity also agrees very well with the literature values. Additionally, the measured thermal conductivity is lower than the amorphous limit (Cahill model), which proves that the diffuson model works better than the Cahill model to describe the thermal conduction mechanism in the amorphous $Nb_2O_5$ thin films. It was also observed that the thermal conductivity does not change significantly with oxygen vacancy concentration. This stable and low thermal conductivity facilitates improved performance for applications such as memristors.

# Experimental

## Sample growth

$Nb_2O_{5-x}$ films were deposited on sapphire substrates by room-temperature reactive DC sputtering using a Nb target in an ambient oxygen environment. Oxygen flow and process pressure were kept constant at 15 sccm and 10 mTorr respectively. Niobium power levels were varied from 25W to 50W to tune the Nb:O ratio, producing films with different oxygen vacancy concentrations.

## XRD characterization

X-ray diffractograms were taken with a Panalytical X'Pert MRD Pro using a double crystal detector and a Cu Kα1 x-ray source. Symmetric 2Θ-ω scans were taken using a ¼ ° slit on the detector and a ¼ ° slit on the source.

## Thermal characterization

The thermal conductivity in this work was measured by TDTR. TDTR is an optical pump and probe method which can be used to measure thermal properties of nanostructured and bulk materials. An ultrafast pump laser beam periodically heats the sample surface while a delayed probe laser beam detects the temperature decay of the sample surface via a change in reflectivity recorded by a photodiode and a lock-in amplifier. The decay curve is fit to a multilayer thermal model to infer unknown thermal properties. More details can be found in the supplementary materials and references[41, 42].

**XPS**

The x-ray photoelectron spectroscopy measurements were all done at room temperature under ultra-high vacuum (~$10^{-7}$ Pa) using a monochromated Al-Kα source (1486.6 eV). Since the samples were charging, Ar ion and electron flood guns were used to neutralize the charging. Binding energy calibration was performed using reference spectra from adventitious carbon (C 1s) and Au foil in electrical contact with the films. More details can be found in the reference.[27]

**Thickness measurements**

Thickness measurements were taken using a Tencor P15 profilometer for films with a thickness larger than 200nm. X-ray reflectivity measurements were taken for films with a thickness less than 200nm. X-ray reflectivity measurements were taken with a Panalytical X'Pert MRD Pro using a double crystal detector and a Cu Kα1 x-ray source. Grazing angle scans were taken using a ¼ slit on the detector and a ¼ slit on the source.

**DFT calculations**

DFT calculations were performed using the VASP program package with the projector augmented wave method (PAW) in the local density approximation.[43-45] The single-crystal $Nb_2O_5$ polytype studied is in the space group 12 (C2/m), with 7 atoms per primitive cell. The lattice structure was relaxed using the primitive unit cell with a convergence of $10^{-8}$ eV for the total energy, and $10^{-7}$ eV/Å for the forces on each atom. The plane-wave energy cutoff was 500 eV. The electronic k-space integration was performed with the tetrahedron method using a Γ-centered 9×9×15 k-mesh. The phonon dispersion is calculated via 3×3×3 primitive cells (189 atoms) by the finite difference method implemented in the PHONOPY package with a Γ-centered 3×3×5 k-mesh.[46]

## Supplementary Information

The supplementary information includes the sensitivity of TDTR measurements, the crystal structure, and XRD results of $Nb_2O_5$ films.

## Conflicts of interest

There are no conflicts to declare.

## Acknowledgements

This work is supported by the U.S. Department of Defense's Multidisciplinary University Research Initiatives (MURI) Program under grant number FOA: N00014-16-R-FO05. Theoretical calculations by T.L.F. and S.T.P. was supported in part by Department of Energy grant DE-FG0209ER46554 and by the McMinn Endowment. Computations at Vanderbilt University and ORNL were performed at the National Energy Research Scientific Computing Center (NERSC), a Department of Energy, Office of Science, User Facility funded through Contract No. DE-AC02-05CH11231. Computations also used the Extreme Science and Engineering Discovery Environment (XSEDE).